\documentclass[aps,
               pra,
               twocolumn,
               nofootinbib,
               floatfix,
               showpacs
              ]{revtex4}

\usepackage{amsmath, amssymb}
\usepackage{graphics}
\usepackage{pstricks}
\usepackage{graphicx}
\usepackage{color}
\usepackage{hyperref}

\renewcommand{\i}{\mathrm{i}}

\renewcommand{\vector}[1]{\mathbf{#1}}

\newcommand{\eq}[1]{(\ref{eq:#1})}
\newcommand{\Eq}[1]{Eq.~(\ref{eq:#1})}

\newcommand{\Fig}[1]{Fig.~\ref{fig:#1}}

\newcommand{\Sect}[1]{Sect.~\ref{sec:#1}}

\newcommand{\subsect}[1]{\ref{sec:#1}}

\begin{document}

% declarations for front matter
\title{Universal scaling at non-thermal fixed points of a two-component Bose gas}

\author{Markus Karl}
%\email{m.karl@thphys.uni-heidelberg.de}
\author{Boris Nowak}
%\email{b.nowak@thphys.uni-heidelberg.de}
\author{Thomas Gasenzer}
%\email{t.gasenzer@uni-heidelberg.de}
\affiliation{Institut f\"ur Theoretische Physik,
             Ruprecht-Karls-Universit\"at Heidelberg,
             Philosophenweg~16,
             69120~Heidelberg, Germany}
\affiliation{ExtreMe Matter Institute EMMI,
             GSI Helmholtzzentrum f\"ur Schwerionenforschung GmbH, 
             Planckstra\ss e~1, 
             64291~Darmstadt, Germany} 

\date{\today}

\begin{abstract}
Quasi-stationary far-from-equilibrium critical states of a two-component Bose gas are studied in two spatial dimensions.
After the system has undergone an initial dynamical instability it approaches a non-thermal fixed point.
At this critical point the structure of the gas is characterised by ensembles of (quasi-)topological defects such as vortices, skyrmions and solitons which give rise to universal power-law behaviour of momentum correlation functions.
The resulting power-law spectra can be interpreted in terms of strong-wave-turbulence cascades driven by particle transport into long-wave-length excitations.
Scaling exponents are determined on both sides of the miscible-immiscible transition controlled by the ratio of the intra-species to inter-species couplings. 
Making use of quantum turbulence methods, we explain the specific values of the exponents from the presence of transient (quasi-)topological defects. 
\end{abstract}

% insert suggested PACS numbers in braces on next line

\maketitle

%== Intro ================================================================
\section{Introduction}
\label{sec:intro}
The concept of universality is used to classify and characterise equilibrium states of matter. 
For example, there are different types of order in a magnetic material separated by a second-order phase transition at which the relevant physical properties become independent of the microscopic details of the system. 
This constitutes a type of universality and allows to characterise an extensive range of different phenomena in terms of just a few classes governed by the same critical properties. 
In view of the intensifying discussion on the dynamics of many-body systems it is a pressing question whether also far away from the thermal limit the character of dynamical evolution can become independent of the microscopic details \cite{Hohenberg1977a}. 
For a closed system this would imply that in approaching a critical configuration the evolution must become independent of the particular initial state the system has started from and critical slowing down in the actual time evolution is observed. 
As a consequence, different types of dynamical evolution could be distinguished by means of universality classes.

Dynamical, self-organised criticality, related to pattern and order formation is a well-explored theme in classical statistical mechanics \cite{Bray1994a}.
The growth of an order parameter in an initially non-ordered system for which the Hamiltonian and the total energy let expect order to develop is commonly described in terms of defect nucleation and coarsening dynamics.
If this process occurs close to an equilibrium configuration, it is described in terms of so-called dynamical critical phenomena, in particular linear-response formulations in the vicinity of equilibrium phase transitions \cite{Hohenberg1977a}.
We emphasize that such descriptions potentially fail when the system is excited sufficiently far from an equilibrium configuration.

Phase-ordering phenomena after domain nucleation are being studied increasingly also in quantum many-body systems, in particular in quantum gases and fluids. 
These systems exhibit, at shorter wave lengths,  quantum nature arising from their microscopic constituents, e.g., cold atoms, but, on the spatial scales of interest for macroscopic ordering phenomena, can be described in terms of classical field theory. 
Related to this, the equilibrium temperature in these situations is sufficiently low such that dynamical critical phenomena at so-called quantum phase transitions can be studied.
Systems are tuned across these transitions by means of, e.g., coupling parameters present in the model.

Described in terms of non-linear field models, the quantum many-body systems bear the possibility of defect formation~\cite{Lee1992a,Nelson2002a}. 
For two-component Bose gases, domain structures~\cite{Timmermans1998a} have been observed in experiments with different hyperfine species of ultracold $^{87}$Rb atoms~\cite{Hall1998a, Hall1998b,Nicklas2011a,Guzman2011a}. 
In addition to these, the system can develop angular-momentum carrying vortices and skyrmions~\cite{Ruostekoski2001a,Kasamatsu2005a}. 
Closely related dynamics appears in spin-1 gases, see Refs.~\cite{Stenger1999a, Miesner1999a, Sadler2006a, Vengalattore2008a, Kawaguchi2010a,Ueda2012a, Fujimoto2012a}.
For discussions of the dynamics near quantum phase transitions see, e.g., Refs.~\cite{Lee2009a,Sabbatini2011a,Polkovnikov2011a}.

Universality and scaling near second-order phase transitions is commonly described by means of the renormalization group (RG).
Critical phenomena arise as fixed points at which the RG flow of effective couplings of a particular model, vanishes.
Varying the microscopic parameters of a model, their subsequent RG flow can pass close by such a fixed point and become critically slowed down.
As a consequence, a wide range of momenta governs the macroscopic correlations of the system,  and scaling behaviour, i.e., power laws in momentum appear.
Here we study the equilibration dynamics of a two-component Bose gas after a quench to a far-from-equilibrium state.
We evaluate correlation functions in view of a corresponding universal evolution in real time.
Our results show that long before the system reaches its final equilibrium state it can approach a non-thermal fixed point \cite{Berges:2008wm} at which correlations exhibit characteristic infrared scaling laws.
Non-thermal fixed points can be derived as stationary scaling solutions of real-time renormalisation-group flows \cite{Berges:2008sr,Gasenzer:2008zz}.
In the system studied here they are found to exist irrespective of whether the two spin components are miscible or the system is in the immiscible regime where the components separate in the low-energy limit.

The type of universal dynamics near a non-thermal fixed point in which we are interested here, has been studied in detail for a single-component Bose gas, in one, two, and three spatial dimensions  \cite{Nowak:2010tm, Nowak:2011sk, Schmidt:2012kw,Schole:2012kt,Nowak:2012gd}.
According to this, the approach of the fixed point is characterised by the creation and dynamical evolution of (quasi-)topological defects \cite{Nowak:2010tm, Nowak:2011sk, Schmidt:2012kw}.
It was shown that equilibration trajectories passing closely by the fixed point lead far away from equilibrium configurations \cite{Schole:2012kt,Nowak:2012gd} and experience critical slowing down.
In the vicinity of the fixed point, power-law and logarithmic decay behaviour in time of the density of vortices was found, corresponding to a growth of the characteristic scale of the pattern \cite{Schole:2012kt,Natu2013a}.
The above studies reveal that the universal dynamics near a non-thermal fixed point is in many respects different from the dynamical critical phenomena known in the vicinity of equilibrium fixed points \cite{Hohenberg1977a,Sieberer2013a}. 

The phenomena predicted at non-thermal fixed points are closely related to classical (wave) turbulence~\cite{Frisch1995a,Zakharov1992a}.
Wave turbulence theory has been extended by field-theoretic methods to quantum systems~\cite{Berges:2008wm, Berges:2008sr, Scheppach:2009wu}.
In the limit of long wavelengths where simple kinetic approximations used to describe weak wave turbulence fail, the non-perturbative methods are used to evaluate high-order correlations.
This allows to predict strong-wave-turbulence scaling laws at non-thermal stationary points of the equations of motion for the correlation functions which could be confirmed by numerical simulations.
Non-thermal fixed points have been demonstrated for relativistic $N$-component scalar theories \cite{Berges:2008wm, Berges:2008sr,Gasenzer:2011by}, relativistic Fermi-Bose systems \cite{Berges2011a}, non-abelian gauge theories \cite{Berges:2008mr} as well as Higgs models \cite{Gasenzer:2013era}.

In this article, we present a numerical analysis of possible non-thermal-fixed-point scalings in a two-component Bose gas in two spatial dimensions tuned to either side  of the miscible-immiscible transition.
We prepare unstable initial conditions to drive the system away from equilibrium and follow the ensuing evolution towards thermalisation. 
Extending upon the results presented in Ref.~\cite{Karl:2013mn} we make use of quantum-turbulence and spin decompositions to trace back the emerging scaling exponents to the presence of various transient domain structures, vortices and skyrmion excitations. 
Similar universal features are expected in the evolution of one-dimensional systems~\cite{Hamner2011a}.

In \Sect{Model} we summarise the main properties of the two-component Bose gas and specify our numerical methodology. 
We furthermore discuss the system in the spin-fluid representation which will be used extensively throughout the paper and use this to exhibit the symmetry properties across the transition between the miscible and immiscible phases.
We finally summarise the analytical results on non-thermal fixed points relevant for the dynamics studied in the following.
\Sect{Simulation} contains our numerical results.
It is divided in detailed discussions of the evolution in the immiscible and immiscible parameter regimes of the model as well as of the dynamics at the transition point.
We present our conclusions in \Sect{summary}.

%
%
%========================================================================================
%== Dynamical Simulations ================================================================
%
\section{The two-component Bose gas}
\label{sec:Model}
In the following section we will present results of quasi-classical simulations of  the dynamics of a dilute two-component Bose gas in two spatial dimensions.
This system is described by the Hamiltonian density ($\hbar = 1$) 
%
%========================================================================================
	\begin{align}
	\label{eq:action}
  \mathcal{H} 
  =  \frac{1}{2m}\nabla\phi_j^{\dagger}\nabla\phi_j + \frac{g}{2}{(\phi_j^{\dagger}\phi_j)}^2 - g(1-\alpha)\phi_1^{\dagger}\phi_1\phi_2^{\dagger}\phi_2\,,
	\end{align}
%=========================================================================================
%
where $m$ is the mass of the atoms,  the sum over the field index $j \in \{1,2\}$ is implied, and $\alpha = g_{12}/g$ the ratio between the inter-species coupling $g_{12}$ and the intra-species interaction constant $g$. 
The coupling $g$ as well as the mass are chosen to be the same for both components.

%==================================================================
\subsection{Equations of motion}
\label{subsec:EoM}
Since we will consider mainly near-degenerate states where the infrared modes of the system are strongly populated we will use quasi-classical simulation techniques~\cite{Blakie2008a, Polkovnikov2010a} to numerically evaluate the time-evolution of correlation functions. 
For this, initial field configurations $\phi_{1,2}(x,t_0)$ are sampled from Gaussian Wigner distributions and then propagated according to the coupled classical equations of motion derived from the Hamiltonian \eq{action}, 
%
%=========================================================================================
      \begin{subequations}
      \label{eq:GP}
	\begin{align}
	\mathrm{i}\partial_t\phi_1 &= -\frac{1}{2}\nabla^2\phi_1 + g(\lvert\phi_1\rvert^2 + \alpha\lvert\phi_2\rvert^2)\phi_1 \label{subeq:GP-1}\,,\\
	\mathrm{i}\partial_t\phi_2 &= -\frac{1}{2}\nabla^2\phi_2 + g(\lvert\phi_2\rvert^2 + \alpha\lvert\phi_1\rvert^2)\phi_2 \label{subeq:GP-2}\,.
	\end{align}
      \end{subequations}
%==================================================================
%                                                                  
At each time during the evolution expectation values can be obtained from ensemble averages over the set of paths sampled. 
We have applied the following rescalings to obtain dimensionless quantities in the system of coupled Gross-Pitaevskii-type equations \eq{GP}, using the lattice constant of the computational grid $a_s$: 
$\phi a_s \to \phi$, $mg \to g$, and $t/(m a_s^2) \to t$. 
Here, $n = (N_1 + N_2)/L^2$ is the mean total particle density on an $N_s \times N_s$ simulation grid of linear size $L = N_s a_s$. 
The healing length derived from the total density reads $\xi = (2mgn)^{-\frac{1}{2}}$ and can be used to map the lattice constant to a value in physical units.
%
%==================================================================
\subsection{Spin-fluid representation}
\label{subsec:SpinRep}
While we will always solve numerically the full set \eq{GP} of equations of motion, it is instructive to concentrate for a moment on the degrees of freedom which describe the relative evolution of the two components. 
Writing the fields in the polar representation $\phi_i = \sqrt{\rho_i}\exp{(\i\varphi_i)}$ the relative degrees of freedom are given by the local phase difference $\theta_r = \varphi_1-\varphi_2$ as well as the local density difference $\rho_1 - \rho_2$. 
With the help of the Pauli matrices $\sigma_a$ the Schwinger representation of angular momentum is defined as $S^a = \phi_j\sigma_{ij}^a\phi_i$ (sum over repeated indices implied). 
This results in a three-component vector of (pseudo-)spin densities ${S^a}$ for $a \in \{x,y,z\}$ which encodes the relative degrees of freedom,
%~\cite{Kasamatsu2005a}, 
%
%=============================================================================
      \begin{subequations}
	\label{eq:SpinRep}
	\begin{align}
	 S^x &= 2\sqrt{\rho_1\rho_2}\cos{\theta_r} \label{subeq:SpinRep-x}\,,\\
	 S^y &= -2\sqrt{\rho_1\rho_2}\sin{\theta_r} \label{subeq:SpinRep-y}\,,\\
	 S^z &= \rho_1 - \rho_2 \label{supeq:SpinRep-z}\,,
	\end{align}
      \end{subequations}
%===============================================================================	
%
where the modulus corresponds to the total density $\lvert\vector{S}\rvert = \rho_1 + \rho_2 \equiv \rho_T$. For convenience, we apply the redefinition $S^a \to \rho_TS^a$ such that 
$\lvert\vector{S}\rvert \equiv 1$. 
Using the above representation, the total energy $E=\int \! \mathrm{d}^2x \, \mathcal{H}$ derived from  \Eq{action} can be written as 
%
%=========================================================================================
	\begin{align}
	\label{eq:energy}
	\nonumber  E = &\int \! \mathrm{d}^2x \, \Bigl[\frac{1}{2}(\nabla\sqrt{\rho_T})^2 + \frac{\rho_T}{8}\nabla S^a  \nabla S^a + \frac{1}{2\rho_T}\vector{j}_T^2 \\
						      &~~+ \frac{g\rho_T^2}{2} -\frac{g\rho_T^2}{4}(1-\alpha)\left[(S^x)^2+(S^y)^2\right]\Bigr]\,,
	\end{align}
%================================================================== 
%
which exhibits the role of the relative degrees of freedom and their coupling to the global ones~\cite{Kasamatsu2005a}. 
The quantity $\vector{j}_T = \rho_1\nabla\varphi_1 + \rho_2\nabla\varphi_2$ is the conserved total particle current associated with the global $U(1)$ phase symmetry of \Eq{action}, i.e., its invariance under a global shift of the total phase $\Theta_T = \varphi_1 + \varphi_2$. 
Thus $\vector{j}_T$ can not be expressed using just the spin densities but contains also the total phase $\Theta_T$. 
We obtain
%
%==================================================================
	\begin{equation}
	\label{eq:totcurr}
	\vector{j}_T = \frac{1}{2}\rho_T\nabla\Theta_T + \frac{S^z\rho_T}{2\left[(S^x)^2+(S^y)^2\right]}(S^y\nabla S^x - S^x\nabla S^y)\,.
	\end{equation}
%==================================================================
%
The representation \eq{energy} of the energy shows that the two-component Bose gas can be equally well described as a spin-carrying fluid with density $\rho_T$ and a (conserved) quasi-particle current $\vector{j}_T$.
For a fluid at rest, i.e., $\rho_T = \mathrm{const}$ and $\vector{j}_T = 0$, the spin system thereby assumes the form of a classical nonlinear sigma model (NL$\sigma$M) with a mass term $g\rho_T^2/4(1-\alpha)\left[(S^x)^2+(S^y)^2\right]$, whereas in general a current $\vector{j}_T \neq 0$ leads to a highly non-trivial coupling between internal and hydrodynamic degrees of freedom. 
%
%
%
%==================================================================
\subsection{Miscible-immiscible transition}
\label{subsec:MiscImmiscTrans}
The two-component Bose gas \eq{action} is well-known to possess two different ground states depending on the value of the parameter $\alpha$~\cite{Timmermans1998a, Kasamatsu2006a}. 
In the immiscible regime, $\alpha > 1$, the inter-species interaction energy overcomes the intra-species interaction.
Hence, in the ground state of the system the spatial overlap of the components is minimised. 
This becomes immediately clear from the expression \eq{energy} since for $\alpha > 1$ the mass term in the NL$\sigma$M,  $g\rho_T^2/4(1-\alpha)\left[(S^x)^2+(S^y)^2\right]$, gives a positive contribution to the total energy for all possible spin configurations, similar to the case of a Heisenberg ferromagnet. 
Thus configurations with $S^z = \pm 1$ are preferred which leads not only to a spin-polarised ground state but also to the formation of domains with oppositely aligned $S^z$ due to the discrete $Z_{2}$ symmetry $S^z \to -S^z$ of the energy \eq{energy}.
On the contrary, in the miscible regime, $\alpha \le 1$, spin configurations with $S^z(x) \equiv 0$ have lowest potential energy. Thus, the zero-temperature ground state is characterised by spatially mixed components with $n_1(x) = n_2(x)$. 
Note that in both regimes the dynamic evolution is constrained by separate exact particle conservation in each component as can be inferred from the structure of interactions in \Eq{action}. 
In this article we will investigate the dynamics of spin and fluid degrees of freedom when driven out of equilibrium, in both, the miscible and immiscible regimes and, in particular, close to the transition point. 
  
%==================================================================
\subsection{Strong wave turbulence}
\label{subsec:SWT}
It is known for the non-equilibrium dynamics of a one-component Bose gas that features of turbulence become manifest during the evolution towards thermal equilibrium. 
In particular, it has been numerically demonstrated in one, two and three dimensions that the single-particle momentum distribution develops universal scaling behaviour during a turbulent stage~\cite{Nowak:2010tm, Nowak:2011sk, Schmidt:2012kw, Schole:2012kt}. 
Here, we are interested in the corresponding observable for the two-component Bose gas  
%
%===============================================================================================================
	\begin{equation}
	\label{eq:sinpartspec}
	n(k,t) = \int \! \mathrm{d}\Omega_k \, \langle \phi_j^{\ast}(\vector{k},t)\phi_j(\vector{k},t) \rangle\,,
	\end{equation}
%===============================================================================================================
%
where $\int \! \mathrm{d}\Omega_k$ denotes the angular average in the two dimensional momentum space. 
In the presence of strong wave turbulence, self similar solutions of the form $n(k) \sim k^{-\zeta}$ are expected. 
In addition to numerical calculations analytic expressions for the scaling exponents $\zeta$ can be found in the context of a generic $O(N)$ symmetric scalar quantum field theory~\cite{Berges:2008wm,Scheppach:2009wu}. 
For the ultraviolet (UV) momentum region, where occupation numbers are low and thus the theory is weakly interacting, kinetic quantum Boltzmann theory is sufficient and results in a weak-wave-turbulence exponent $\zeta^\mathrm{UV} = d = 2$~\cite{Zakharov1992a,Nazarenko2011a}. 
In the infrared (IR) region, high occupation numbers render the theory strongly interacting and therefore non-perturbative methods have to be used. 
An infinite resummation of a certain class of Feynman diagrams was shown to yield a scaling exponent in the IR,
%
%===============================================================================================================
	\begin{equation}
	\label{eq:scalingexp}
	\zeta^\mathrm{IR} = d + 2 = 4\,,
	\end{equation}
%===============================================================================================================
%
which corresponds to an inverse cascade of particles from intermediate momenta towards the IR~\cite{Scheppach:2009wu}. 
Finally, note that also the thermalised system is characterised by a scaling solution for $n(k)$ with an exponent $\zeta^\mathrm{th} = 2$, corresponding to a Rayleigh-Jeans momentum distribution.

%==================================================================
%==================================================================
\section{Dynamical simulations}
\label{sec:Simulation}
In the following, we study the time evolution of a two-component Bose gas in the miscible and immiscible regimes, starting from on-average unpolarised nonequilibrium initial states. 
The parameters of our simulation are chosen such that the final state is close to the groundstate of the system. Making use of dynamical instabilities, we drive the system far from equilibrium 
and study the transient properties of ensemble averaged correlation functions. A special focus is set on the relation between scaling properties and their microscopic origin.

% 
%===============================================================================================================
\subsection{Dynamics in the immiscible regime}
\label{subsec:ImmiscRegime}
%
%-------------------------------------------------------------------------------------------------
\subsubsection{Formation of polarisation patterns and defects}
%  
%===============================================================================================================
    \begin{figure}[!t]
    \includegraphics[width=0.50\textwidth]{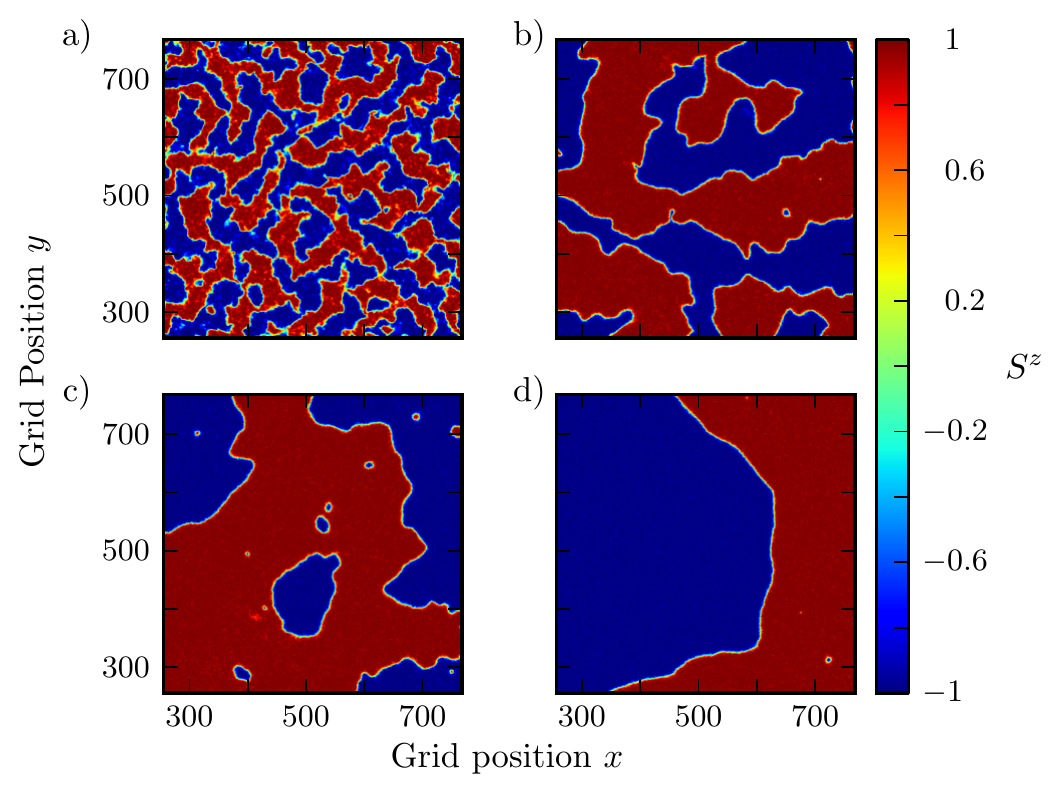}
    \includegraphics[width=0.50\textwidth]{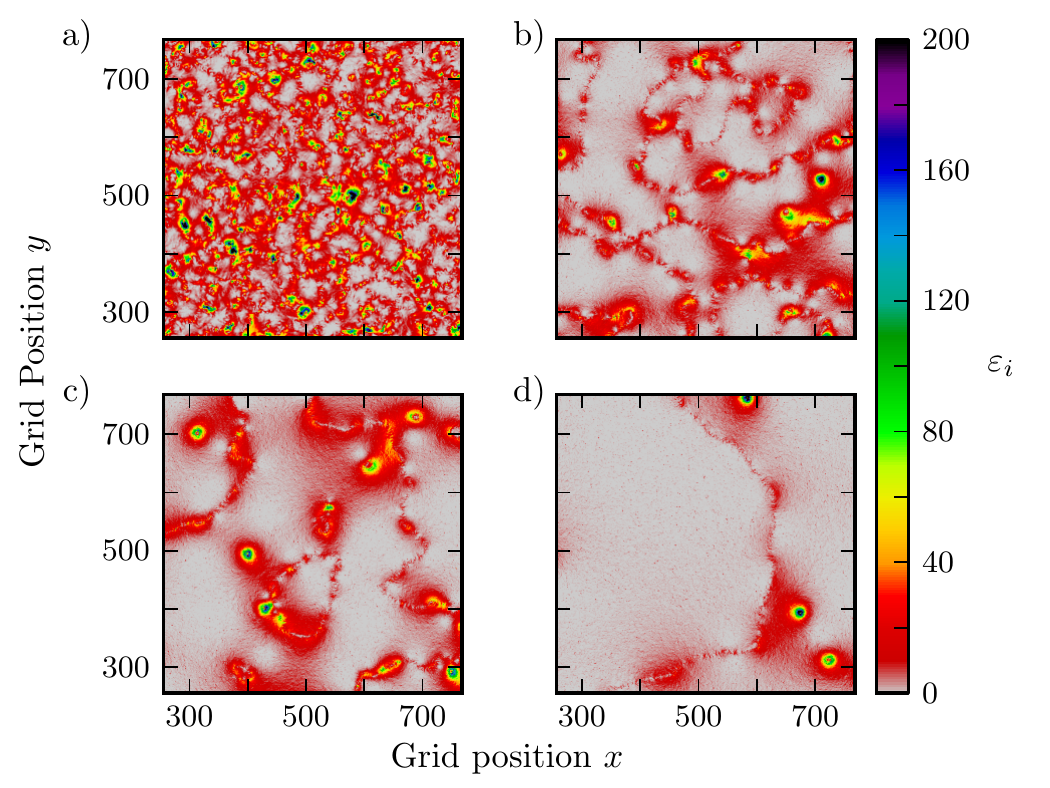}
    \caption{(Colour online) The figure shows the time evolution of the spin density $S^z(\vector{x})$ (upper half) and of the incompressible kinetic energy density 
$\epsilon_i = \frac{1}{2}\lvert\vector{w}_i(\vector{x})\rvert^2$ (lower half) in the immiscible regime. Numerical parameters are $N_s = 1024$, $N_1 = N_2 = 3.2\cdot 10^9$, $g = 3\cdot 10^{-5}$ and $\alpha = 2$. 
Panels (a) to (d) correspond to snapshots of a single run at different grid times $t = 500, 5000, 10000, 100000$. 
Note that only a $512^2$ subsection of the computational grid is displayed in order to highlight the emerging structures.
Quickly after the start of the simulation domains with oppositely oriented spin emerge as an isotropic pattern, and incompressible energy is created along the domain borders, see panels (a). 
During the ensuing evolution towards the equilibrium state these domains merge until the system reaches a state in which only two large domains exist, cf.~panels (d).
On top of the coarse domain structure point-like domains become visible during the intermediate stage of the evolution which are revealed to be skyrmionic defects by the accompanying pattern of incompressible energy. 
See the main text for further details.
}
    \label{fig:immdens}
    \end{figure}
%===============================================================================================================
%===============================================================================================================
    \begin{figure}[!t]
    \includegraphics[width=0.49\textwidth]{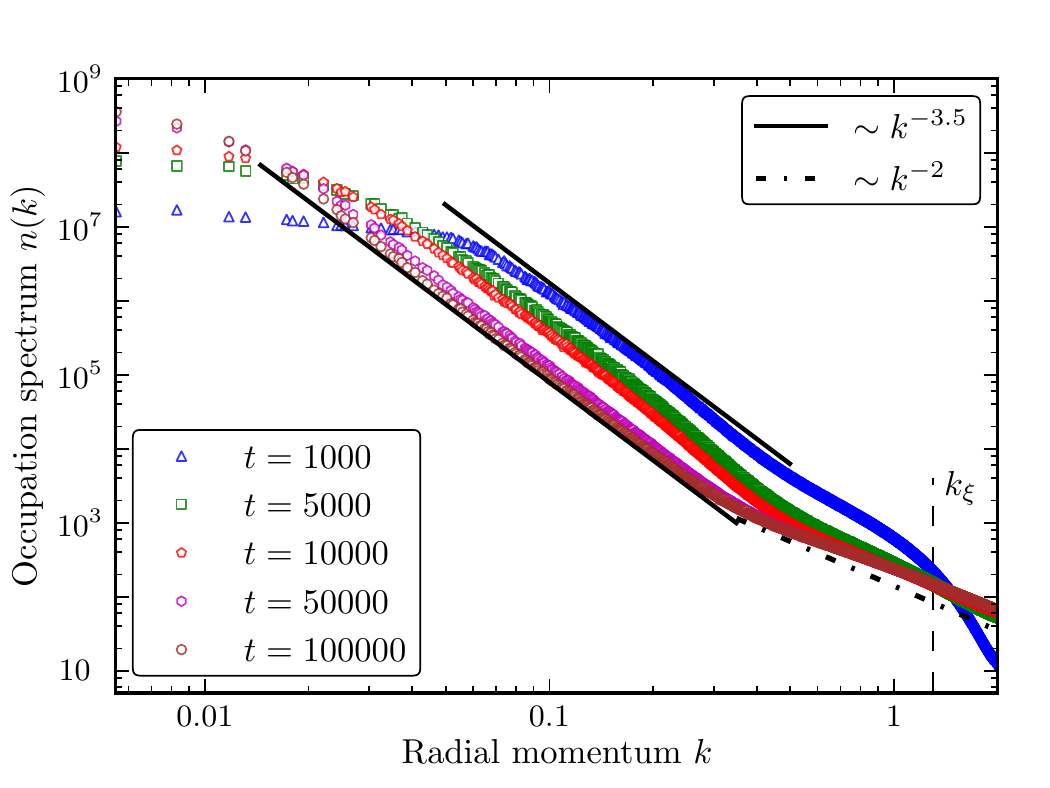}
    \caption{(Colour online) The single-particle momentum distribution as defined in \Eq{sinpartspec} at different  time steps of the evolution, in dimensionless units defined in the main text. 
    Numerical parameters correspond to \Fig{immdens} and the spectra are averaged over 100 runs. 
    After the onset of domain formation a scaling regime with $n \sim k^{-3.5}$ forms at intermediate momenta. 
    At a time $t = 5000$ this region begins to move towards the IR while a new scaling region with $n \sim k^{-2}$ develops in the UV. 
    The latter signals the onset of thermalisation of the UV modes. 
    At late times a stable bimodal power law behaviour has formed with a scaling exponent $\zeta = 2$ in the UV and an IR scaling exponent $\zeta \simeq 3.5$.
    }
    \label{fig:immspec}
    \end{figure}
%===============================================================================================================
For the simulations in the immiscible regime we choose $\alpha = 2$ and initial unpolarised configurations, $\langle S^z(\vector x, t=0)\rangle = 0$, and add quantum noise to each mode.
Specifically, we take only the zero mode of both complex fields to be initially occupied with a macroscopic number of particles $\phi_i(\vector k, t=0) = \sqrt{N_i}\delta_{\vector k,0} + c_{\vector{k}}$, where the $c_{\vector{k}}$ are Gaussian distributed complex numbers with random phase, $\langle c \rangle = 0$ and $\langle c^{\ast}c \rangle = {1}/{2}$. 
We choose the number of particles in each component to be equal on average, $N_1 = N_2$. 
Initial configurations of such type are far from the energetically preferred state and therefore lead to instabilities that drive the system dynamically towards spatial demixing~\cite{Timmermans1998a, Kasamatsu2006a}. 
A typical time evolution of one realisation of the described initial configurations is depicted in \Fig{immdens}. 
We follow the evolution until the spin system resembles its energetic ground state configuration, implying the total energy to be close to the ground-state energy. 
This means that, consistent with particle conservation in each component, only two domains exist which are separated by a nearly straight domain wall. 
Compared to the full simulation time the demixing process occurs on a much faster time scale which is given by the energy of the fastest growing unstable mode, 
$t_I \simeq 1/\omega_I$. The main part of the evolution is dominated by merging of domains which leads to a coarse-graining of the domain structure.\\  
%

%-------------------------------------------------------------------------------------------------
\subsubsection{Scaling and non-thermal fixed point}
To show the relation of the above results to the approach of a non-thermal fixed point, we consider the evolution in terms of the single-particle spectrum \Eq{sinpartspec}, see \Fig{immspec}. 
We find that already at an early stage of the evolution a power-law distribution has developed in the IR regime, $k\lesssim0.4$. 
This scaling behaviour terminates at a decreasing infrared cutoff scale $\pi/L_\mathrm{D}$ given by the mean domain size $L_\mathrm{D}$.
The cut-off at the UV end at $\pi/\xi_s$ is found to be approximately set by the width of the domain walls, i.e., the spin healing length $\xi_s =\xi (2 / \lvert1-\alpha\rvert)^{{1}/{2}}$. 
In between these scales, a scaling region with an exponent of $\zeta \simeq 3.5$ emerges. 
Consistent with the observed spatial growth of the domains of uniform spin polarisation, the IR limit of the scaling regime shifts in time towards lower momenta.
In the `far' UV, quasi-classical thermalisation of the distribution sets in which is signalled by the appearance of the Rayleigh-Jeans scaling exponent $\zeta = 2$. 
During the whole time evolution, the IR scaling exponent stays at $\zeta \simeq 3.5$.
This exponent stands in contrast with the strong-wave-turbulence prediction for a non-relativistic field theory in $d=2$ dimensions which is $\zeta^\mathrm{IR} = 4$, cf.~\Eq{scalingexp} and Ref.~\cite{Scheppach:2009wu}.\\
%
%===============================================================================================================
    \begin{figure}[!t]
    \includegraphics[width=0.49\textwidth]{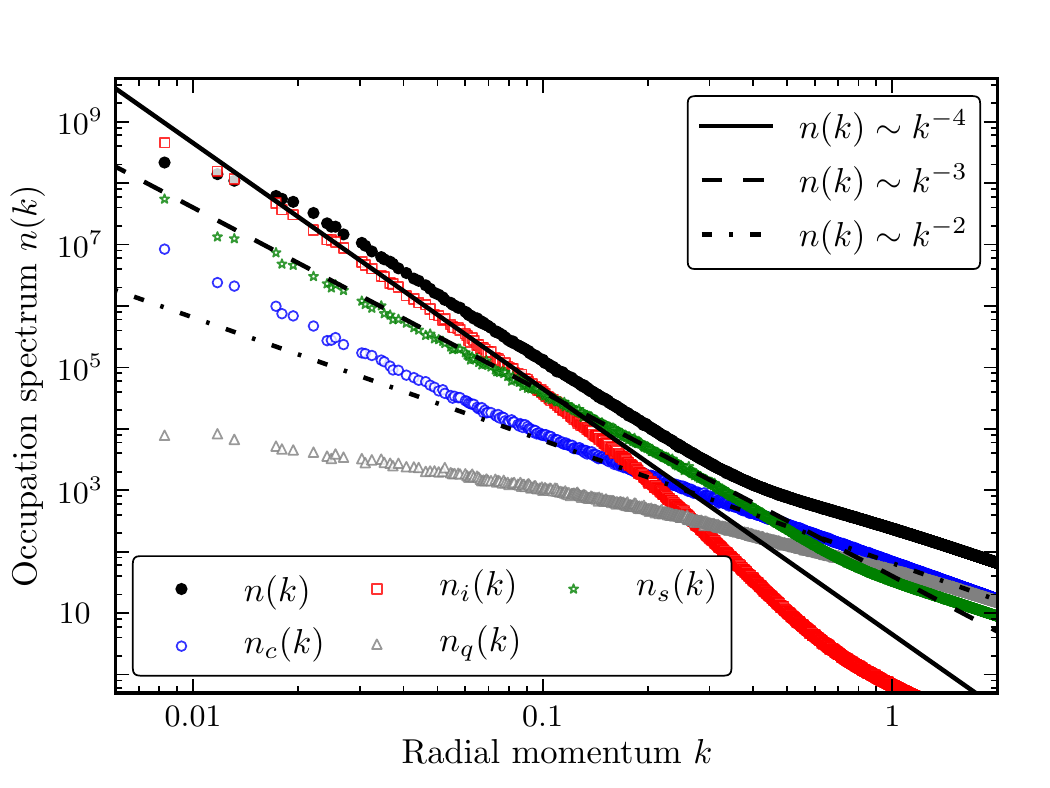}
    \caption{(Colour online) The decomposition of the occupation spectrum as defined in \Eq{radenergyspec}, in comparison to the full spectrum at grid time $t = 5\cdot10^{4}$, with numerical parameters corresponding to \Fig{immdens}. Averages are taken over 100 runs. The black dots show the full spectrum which is also displayed in \Fig{immspec} while open coloured symbols refer 
to the different parts of the decomposition, the incompressible and compressible parts of the classical hydrodynamic kinetic energy $n_i$ (red squares) and $n_c$ (blue circles), the quantum pressure 
part $n_q$ (grey triangles) and the spin pressure part $n_s$ (green stars). 
The IR scaling behaviour $n\sim k^{-\zeta}$ is dominated by incompressible flow generated by vortex excitations with a scaling exponent $\zeta = 4$.
Spin excitations form domain walls which also give an important contribution in the same momentum regime but with a scaling exponent $\zeta = 3$, such that the sum appears to follow a scaling law with $\zeta \simeq 3.5$. 
See main text for further details.} 
    \label{fig:immspec_decomp}
    \end{figure}
%===============================================================================================================
%
%-------------------------------------------------------------------------------------------------
\subsubsection{Hydrodynamic and spin-fluid decomposition}
In order to gain further insight into the interplay between the domain structure and other excitations we decompose the energy along the lines introduced in Refs.~\cite{Nore1997a, Nore1997b}, see also \cite{Nowak:2011sk}.
According to the representation \eq{energy} of the energy, the kinetic part of the original Gross-Pitaevskii model \eq{action} can be written in the following way
%
%===============================================================================================================
	\begin{align}
	\label{eq:energydec}
	E_{\mathrm{kin}} 
%	&= \frac{1}{2}\int \! \mathrm{d}^2x \, \nabla \phi_j \nabla \phi_j
%	\nonumber \\ 
	&= \frac{1}{2}\int \! \mathrm{d}^2x \, \Bigl[\lvert\nabla\sqrt{\rho_T}\rvert^2 + \frac{\rho_T}{4}\nabla S^a  \nabla S^a + \lvert\vector{w}\rvert^2\Bigr] \,.
	\end{align}
%===============================================================================================================             
%
The velocity field $\vector{w}$ is defined via the total particle current $\sqrt{\rho_T}\vector{w} = \vector{j}_T$, similar to the convenient choice for the one-component case. 
In this decomposition, the first and the last term are the quantum-pressure and the classical hydrodynamic components, respectively. 
In contrast to a single-component fluid, the second term of \Eq{energydec} adds a pressure-like contribution to the kinetic energy which is produced by internal excitations only. 
In addition, the velocity field $\vector{w}$ and thus the corresponding part of the kinetic energy can be decomposed in a compressible and an incompressible part, $\vector{w} = \vector{w}_c + \vector{w}_i$, with $\nabla \times \vector{w}_c = 0$ and $\nabla \cdot \vector{w}_i = 0$. 
In such a decomposition, the effects of wave-like and vortical excitations show up in different parts. 
For sound waves, the compressible part becomes the important one while vortices appear in the incompressible part of the decomposition~\cite{Nore1997a, Nore1997b}. 
Based on \Eq{energydec}, radial energy spectra in momentum space which correspond to the respective parts of the energy can be defined,
%
%===============================================================================================================
	\begin{subequations}
	\label{eq:radenergyspec}
	  \begin{align}
	      E_{\delta}(k) &= \frac{1}{2} \int \! \mathrm{d}\Omega_k \, \langle \lvert\vector{w}_{\delta}(k)\rvert^2 \rangle,~~~\delta \in \{q,c,i\}, \label{subeq:rad-1} \, \\
	      E_{s}(k) &= \frac{1}{2} \int \! \mathrm{d}\Omega_k \, \langle \vector{w}_{s}^a(k) \cdot \vector{w}_{s}^a(k)  \rangle \,. \label{subeq:rad-2}  
	  \end{align}
	\end{subequations}
%=============================================================================================================== 
%
Here we have introduced additional velocities $\vector{w}_q = \nabla\sqrt{\rho_T}$ and $\vector{w}_s^a = \sqrt{\rho_T}/2\nabla S^a$ for the sake of a closed representation. 
Finally, the energy spectra can be converted to occupation-number spectra in the sense of \Eq{sinpartspec} by multiplication with a factor $k^{-2}$, $n_{\delta}(k) = k^{-2}E_{\delta}(k)~~~\delta \in \{q,c,i,s\}$~\cite{Nowak:2011sk}. 
A decomposition of such type in comparison to the full occupation spectrum at a late stage of the evolution ($t=50000$) is shown in \Fig{immspec_decomp} corresponding to the evolution depicted in \Fig{immspec}. 
We find that compressible and pressure excitations dominate the occupation spectrum in the UV region following a thermal distribution $n(k) \sim k^{-2}$ while they give a negligible contribution except for the spin pressure component $n_s$ towards lower momenta. 

%-------------------------------------------------------------------------------------------------
\subsubsection{Ensemble of vortex and skyrmion defects}
The main contribution to the spectrum in the IR momentum regime is provided by incompressible excitations $n_i$ which is characteristic for quantum turbulence whereas the spin excitations $n_s$ overtake in a regime of intermediate momenta. 
Thereby, the incompressible spectrum shows an $n_i \sim k^{-4}$ scaling over approximately one decade for low momenta which is generated by coherent vortical flows $\vector{w}_i$ around topological defects. 
To elaborate on this, the lower part of \Fig{immdens} shows incompressible energy densities $\varepsilon_i(\mathbf{x}) = \lvert\vector{w}_i(\mathbf{x})\rvert^2/2$ for a single run. 
Here we observe that for later stages of the evolution incompressible energy is mainly distributed around isolated points which correspond to either vortices in one component or vortices filled with the other component, the latter being a variant of a so-called skyrmion \cite{Ruostekoski2001a,Kasamatsu2005a}. 
These defects get created during the merging process of the domains and, persisting due to their topological nature, give the main contribution to the incompressible excitation spectrum with the scaling exponent $\zeta = 4$. 
For very late stages of the time evolution it can be seen that those skyrmionic defects can also be created or annihilated by excitations of the domain walls and thus the point defects are present during the whole simulation. 
The spin excitation spectrum, on the other hand, displays a scaling of $n_s \sim k^{-3}$ in the low- and intermediate-momentum regimes due to the existence of domain walls. 

%-------------------------------------------------------------------------------------------------
\subsubsection{Solitary walls}
Under the assumption $\rho_T = \mathrm{const}$, the spectrum $n_s$ can be related to the Fourier transform of the correlation function of the spin order parameter 
$\mathcal{S} = \int \! \mathrm{d}\Omega_k \, \langle S^a(-\vector{k})S^a(\vector{k})\rangle$,
	\begin{align}
	\label{eq:strucfunc}
	\nonumber k^{2}n_s &= \frac{\rho_T}{2} \int \! \mathrm{d}\Omega_k \, \langle \mathcal{F}(\nabla S^a)^{\ast} \mathcal{F}(\nabla S^a) \rangle \\
			    &= \frac{\rho_T}{2} k^2 \int \! \mathrm{d}\Omega_k \, \langle S^a(-\vector{k}) S^a(\vector{k}) \rangle,  
	\end{align}     
with $\mathcal{F}$ denoting the Fourier transform, and therefore $n_s = {\rho_T}\mathcal{S}/2$. Hence, in the regime ${\pi}/{L_D} \ll k \ll {\pi}/{\xi_s}$ where the 
scaling behaviour is dominated by a single domain wall $n_s \sim \mathcal{S} \sim k^{-3}$ follows, e.g., from an ensemble of configurations with $S^z(\vector{x}) = 1 - 2\Theta([\vector{x}-\vector{x}_{0}]\vector{e}_{\perp})$, $\langle S^x\rangle \equiv \langle S^y\rangle \equiv 0$, with random position $\vector{x}_{0}$ and orientation of vector $\vector{e}_{\perp}$ normal to the wall. 
This feature is similar to the scaling induced by solitons in one-dimensional Bose gases, where a phase jump occurs in the bosonic field $\phi(x) = \pm(1 - 2\Theta(x))$ and induces a scaling $n_{1D}\sim k^{-2}$, see Ref. \cite{Schmidt:2012kw}. 

Since in \Fig{immspec_decomp} the two contributions $n_s$ and $n_i$ are of comparable magnitude in an intermediate momentum range the sum of all contributions, which gives the full spectrum, appears to follow the scaling law $n \sim k^{-3.5}$ in the IR. 
However, the momentum range in our simulations is limited by an IR cut-off determined by the grid size. 
We expect that, on a much larger grid, the two superimposed scaling powers to separate such that in the limit $k \to 0$ a clear exponent $\zeta = 4$ arises while in an intermediate momentum region  $\zeta = 3$ dominates.
%
%
%
%
%
%===============================================================================================================
\subsection{Dynamics in the miscible regime}
\label{sec:MiscRegime}
%
%-------------------------------------------------------------------------------------------------
\subsubsection{Initial condition and counter-superflow instability}
For the simulations in the miscible regime we used initial field configurations which are spatially homogeneous in the density of both components and unpolarised, $\langle S^z(\vector x, t=0)\rangle \equiv 0$ except for quantum noise. 
In contrast to the immiscible regime, configurations of such type are close to the energetically favoured state of the spin system.
Therefore a special mechanism is needed to drive the spin system towards a non-equilibrium state. 
For this we exploit the so-called counter-superflow instability (CSI) which is a hydrodynamic instability known to exist in miscible ultracold two-component gases~\cite{Takeuchi2010a, Takeuchi2011a}. 
Both gas components are chosen to be spatially homogenenous in density but possess constant counter-directed current fields.
Hence, each realisation has the form $\phi_{j}(x) = \sqrt{\rho_j}\exp\{\i \vector{v}_j \cdot \vector{x}\}$, with $\vector{e}_{x}$-directed constant velocity fields $\vector{v}_1 = -\vector{v}_2 = {v}\vector{e}_x/2 $. 
A Bogoliubov--de Gennes analysis of this initial state reveals a critical instability for certain choices of the counterflow velocity $v$. 
In particular, if $v$ exceeds a critical value, $v > \xi^{-1}\sqrt{1-\alpha}$, unstable momentum modes exist in the system and momentum exchange between the two superfluid components is possible ~\cite{Takeuchi2011a}. 
For the dynamical simulations we choose the initial states to be 
\begin{align}
 \phi_1(\vector k, t=0) 
 &= \sqrt{N_1}\delta_{k_x,{v}/{2}}\delta_{k_y,0} + c_{\vector{k}},
 \nonumber\\
\phi_2(\vector k, t=0) 
&= \sqrt{N_2}\delta_{k_x,-{v}/{2}}\delta_{k_y,0} + c_{\vector{k}}. 
\label{eq:initdist}
\end{align}     
As before, there are random complex numbers added to each momentum mode such that $\langle c \rangle = 0$ and $\langle c^{\ast}c \rangle = {1}/{2}$ to generate an appropriate ensemble of initial field configurations. 
The relative interaction strength is set to $\alpha = 0.8$ and the initial counterflow velocity is set to $v = 0.70\xi^{-1}$ such that $v > v_{\mathrm{crit}} = \sqrt{0.2}\xi^{-1}$.
% 
%

%-------------------------------------------------------------------------------------------------
\subsubsection{Pattern and defect formation}
\begin{figure}[!t]
    \includegraphics[width=0.49\textwidth]{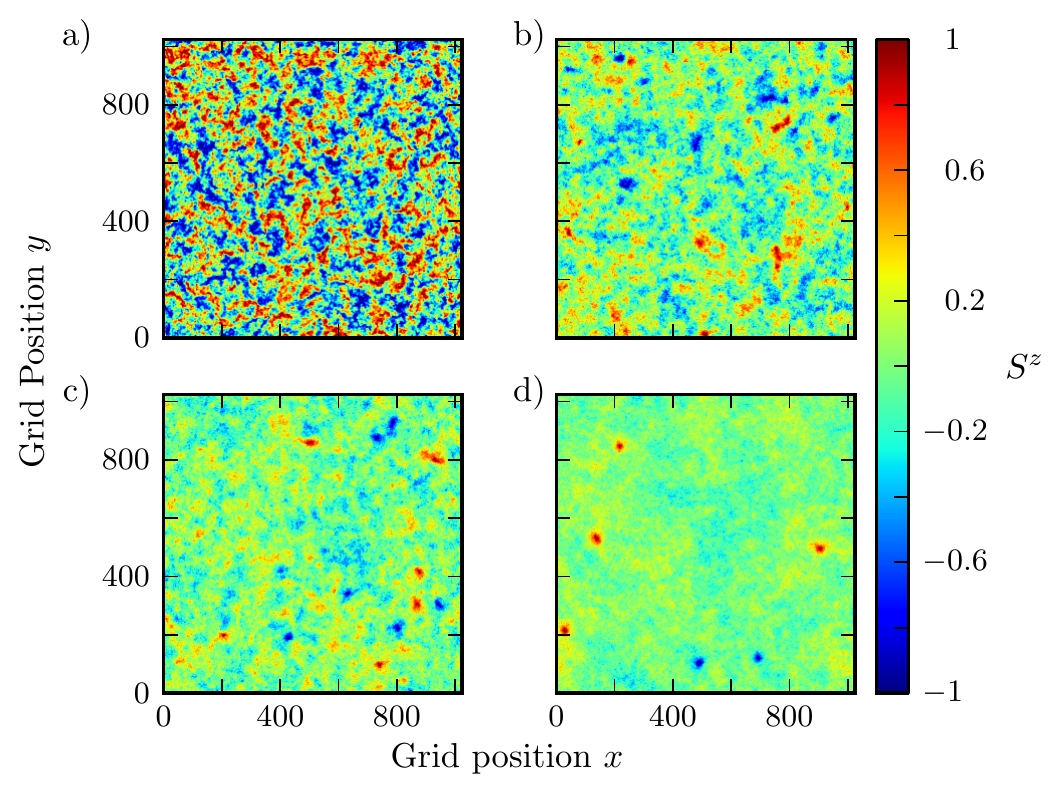}
\includegraphics[width=0.49\textwidth]{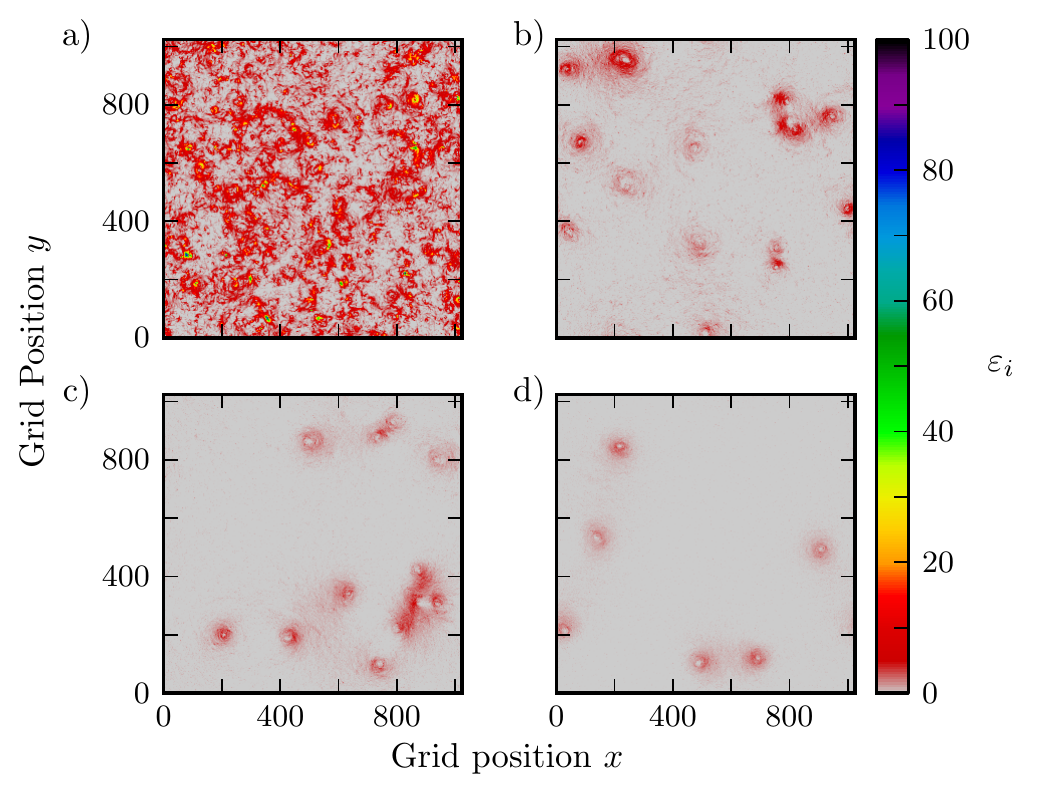}
    \caption{(Colour online) The time evolution of the spin density distribution $S^z(\vector{x})$ (upper half) and of the incompressible kinetic energy density 
$\epsilon_i = \lvert\vector{w}_i(\vector{x})\rvert^2/2$ (lower half) in the miscible regime. 
Numerical parameters are $N_s = 1024$, $N_1 = N_2 = 3.2\cdot 10^9$, $g = 1\cdot 10^{-5}$, and $\alpha = 0.8$, with $v=\xi^{-1}$. 
Panels (a) to (d) correspond to snapshots of a single run at different grid times $t = 5000, 50000, 100000, 500000$. 
After the onset of the counter-superflow-instability, isotropic $S^z$-polarised spin configurations emerge at intermediate times which are accompanied by uniform distributions of incompressible energy $\epsilon_i$ (see Panel (a)). 
During the thermalisation process, the polarisation decays in most of the spatial grid, thereby revealing the existence of vortex--anti-vortex pairs in 
both components which have $S^z \neq 0$ at the centre of their cores. 
These point defects persist during the whole run time of the simulation and dominate the dynamical features during the late stage (Panel (d)). 
Consequently, the incompressible energy concentrates around vortical excitations during the later stage of the evolution (Panels (b--d)), passing from classical to quantum turbulence.}
    \label{fig:mdens}
    \end{figure}

\begin{figure}[!t]
    \includegraphics[width=0.49\textwidth]{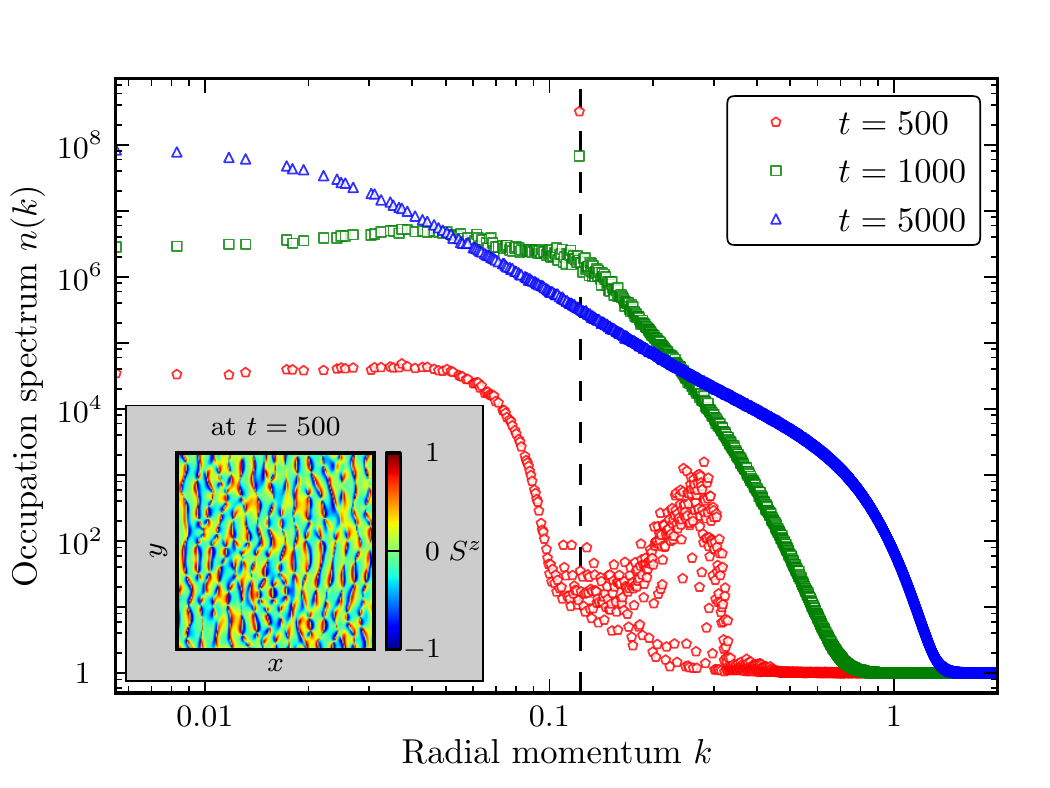}
    \caption{(Colour online) 
    The single-particle momentum distribution as defined in \Eq{sinpartspec}, at different grid times during the onset stage of the counter-superflow-instability (CSI). 
The numerical parameters correspond to \Fig{mdens}, and the spectra are averaged over 100 runs. 
Initially only modes $k_v$ corresponding to the counterflow velocity are macroscopically occupied (marked by the vertical dashed line).
The CSI triggers the growth of certain unstable modes which then act as a source positioned at an intermediate momentum scale. 
This initiates a typical cascading process in momentum space, causing the transport of predominantly particles to the lower momentum modes and of energy to the higher momentum modes. 
The inset shows a snapshot of a typical spin configuration $S^z(\vector{x})$ at  $t = 500$. 
One finds that the CSI results in planar spin waves propagating in the direction of the initial counterflow which we chose pointing into the $x$-direction.}
    \label{fig:mspecasy_onset}
    \end{figure}

In the early stages of the time evolution the dynamical onset of the discussed instability generates planar spin waves, i.e., kinks in the $S^z$ field which propagate along the counterflow direction. 
A typical spin configuration for this early stage of the evolution is depicted in the inset of \Fig{mspecasy_onset}. 
Since these (pseudo-)kink solutions are energetically unfavourable in the miscible regime they decay very fast, thereby driving the system through a phase of turbulent evolution. 
\Fig{mdens}, upper set, Panel (a) shows that after the initial evolution stage the spin system develops an isotropically $S^z$-polarised form that is similar to the domain structure which builds up in the immiscible regime. 
Instead of a coarsening, however, the polarised structure decays in the following evolution towards thermal equilibrium (see Panels (b) to (d) in the upper part of \Fig{mdens}) due to the lack of topologically protected kink solutions in the miscible regime. 
During the initial evolution we observe also vortex--anti-vortex pairs in both components being created by the instability. 
These defects survive for a much longer timescale than the intermediate isotropic spin polarisation. 
The late stage of the evolution is therefore dominated by vortex pair dynamics, with algebraic decay in time as found in simulations of one-component gases~\cite{Nowak:2011sk,Schole:2012kt}. 
Panel (a) in the lower part of \Fig{mdens} shows that the intermediate spin configurations support also isotropic and smooth distributions of incompressible kinetic energy $\epsilon_i = \lvert\vector{w}_i(\vector{x})\rvert^2/2$, similar to a classical turbulent flow. 
With the decay of the spin polarisation also the incompressible energy vanishes in the major parts of the simulation space and instead concentrates around the persistent vortical excitations (see Panels (b) and (c) in the lower part of \Fig{mdens}). 
Consequently, in the late stage of the evolution incompressible energy is distributed only in the vicinity of point defects, i.e., vortices, as is expected for quantum turbulent flows (see panel (d) in the lower part of \Fig{mdens}).\\  
%===============================================================================================================
%===============================================================================================================
%===============================================================================================================
%===============================================================================================================

%-------------------------------------------------------------------------------------------------
\subsubsection{Turbulent scaling}
\begin{figure}[!t]
    \includegraphics[width=0.49\textwidth]{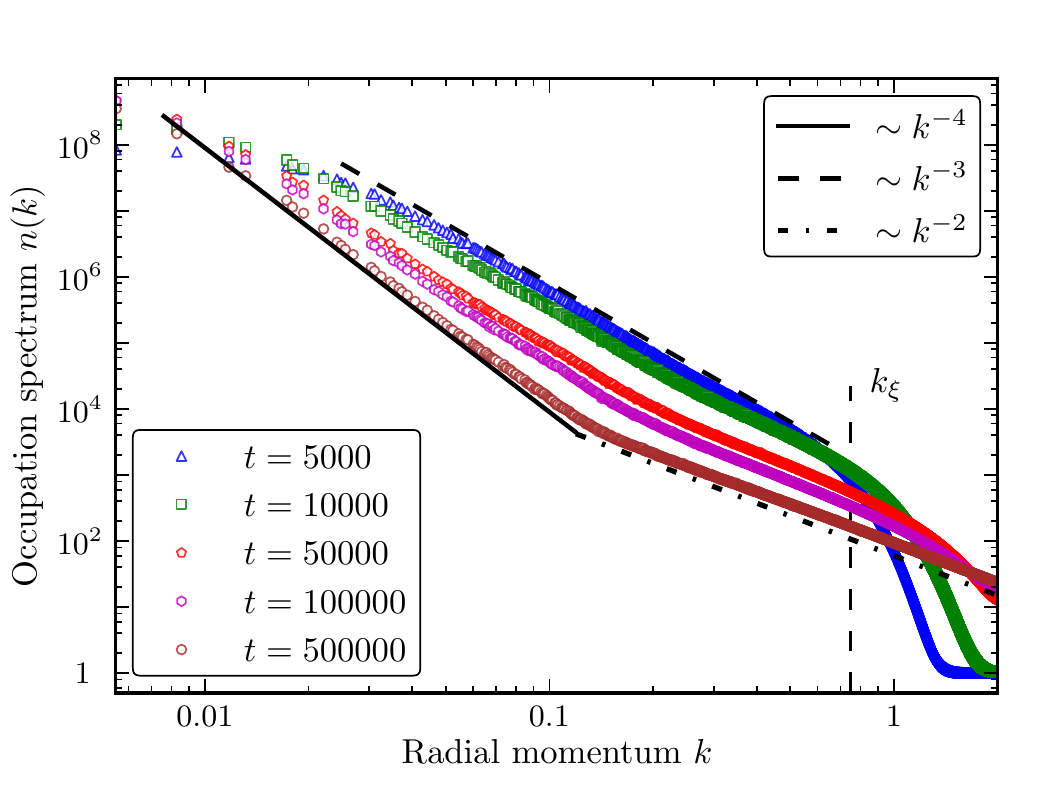}
    \caption{(Colour online) The single-particle momentum distribution as defined in \Eq{sinpartspec} at different grid times, in the miscible regime. 
    Numerical parameters correspond to those in \Fig{mdens}, and the spectra are averaged over 100 runs. 
    With vanishing $S^z$ polarisation also the scaling distribution $n(k) \sim k^{-3}$, which is seen at intermediate times over a wide momentum interval changes 
towards a bimodal form. 
While high momentum modes thermalise, forming a Rayleigh-Jeans distribution $n(k) \sim k^{-2}$, the IR regime is dominated by vortical excitations, implying 
$n(k) \sim k^{-4}$.}
    \label{fig:mspecasym}
    \end{figure}

Also in the miscible regime, signals of turbulent evolution should become manifest in scaling solutions for the single-particle momentum spectrum \eq{sinpartspec} as well as of its different hydrodynamic contributions according to the decomposition of the radial kinetic energy density \eq{radenergyspec}. 
As is shown in \Fig{mspecasy_onset}, initially only modes $k_v$ corresponding to the counterflow velocity are macroscopically occupied (marked by the vertical dashed line). 
In the early stage of the time evolution these modes feed two processes simultaneously: 
Due to the CSI unstable modes grow exponentially in the low-momentum regime. 
Scattering between particles leads to power-law growth of momentum modes for $k>k_v$. 
Ultimately, this results in the build-up of an inverse cascade which transports particles to the lower momentum modes. 
At grid time $t = 5000$ (last timestep in \Fig{mspecasy_onset} and first in \Fig{mspecasym}) a scaling  with $\zeta = 3$ appears, corresponding to randomly distributed (pseudo-)kinks in the densities and thus in $S^z$ (Panel (a) in \Fig{mdens}). 
\Fig{mspecasym} depicts the further time evolution of the single-particle spectrum. 
After the intermediate, kink-dominated stage the spectrum develops a bimodal form with $n(k) \sim k^{-4}$ in the IR and $n(k) \sim k^{-2}$ in the UV.
This scaling is well known from vortex and vortex pair dynamics in a two dimensional one-component gas~\cite{Nowak:2011sk,Schole:2012kt}. 
Consistent with the field theoretical predictions and the existence of a non-thermal fixed point, the system stays for the majority of the simulated time in this vortex-dominated stage.\\   
%===============================================================================================================
%===============================================================================================================
%===============================================================================================================
    \begin{figure}[!t]
    \includegraphics[width=0.49\textwidth]{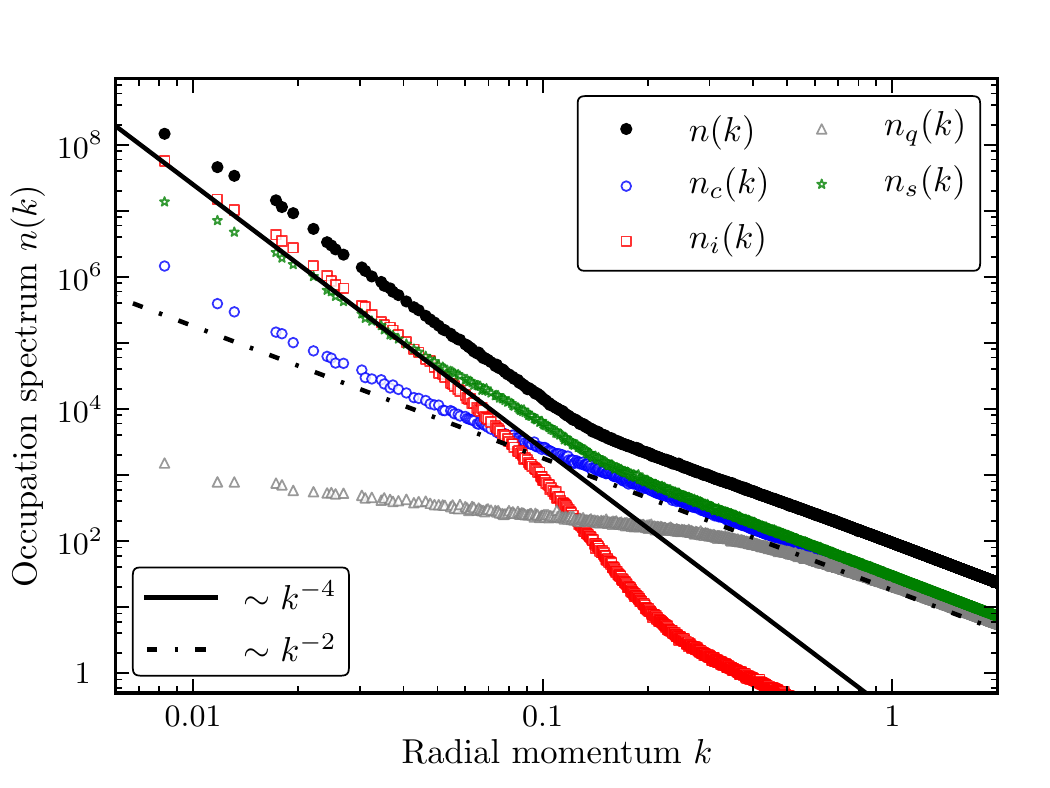}
    \caption{(Colour online) Decomposition of the momentum spectrum as defined in \Eq{radenergyspec}, in comparison with the full spectrum at grid time $t = 500000$.
     Numerical parameters correspond to those in \Fig{mdens}. 
     Averages are taken over 100 runs. 
     The black dots show the full spectrum also displayed in \Fig{mspecasym}.
     Open coloured symbols refer to the different parts of the decomposition:
     The incompressible and compressible parts of the classical hydrodynamic kinetic energy $n_i$ (red squares) and $n_c$ (blue circles), 
the quantum pressure part $n_q$ (grey triangles) and the spin pressure part $n_s$ (green stars). 
Other than in the immiscible regime the spin pressure contribution $n_s(k)$ develops a scaling $n(k) \sim k^{-4}$ in the IR region which adds up with the incompressible energy distribution $n_i(k)$ to give an IR scaling exponent $\zeta = 4$.
See main text for further details.}
    \label{fig:mspecasy_decomp}
    \end{figure}
%===============================================================================================================
%
%===============================================================================================================
    \begin{figure}[!t]
    \includegraphics[width=0.49\textwidth]{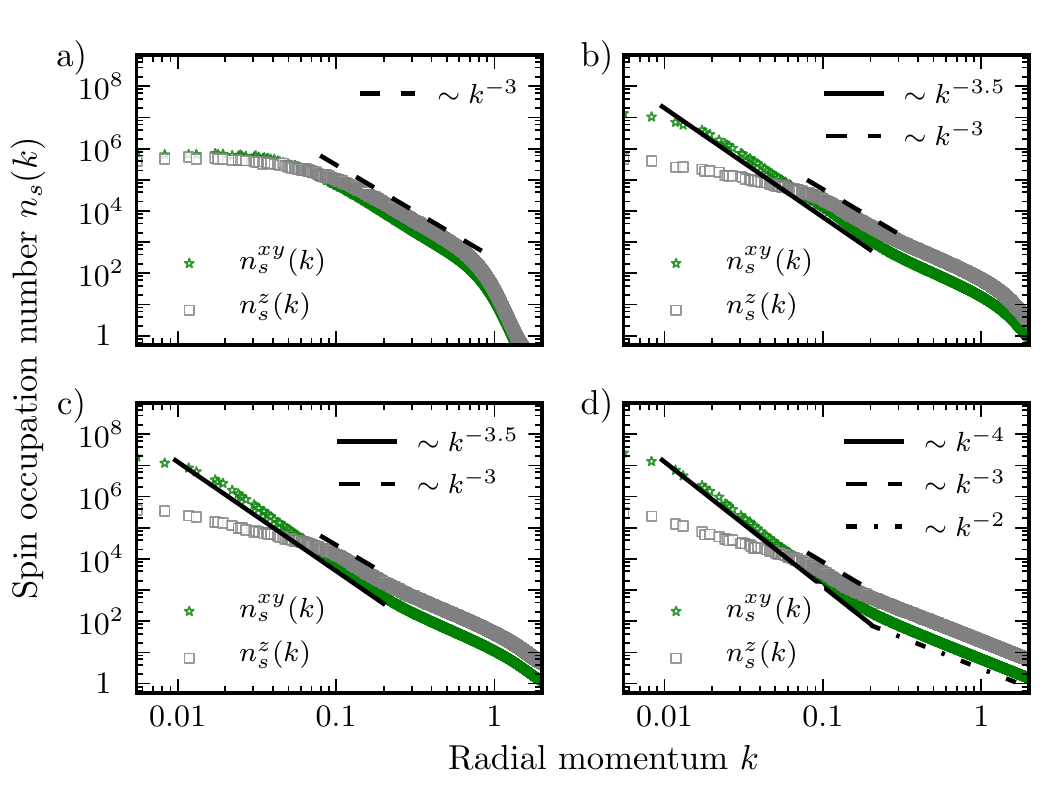}
    \caption{(Colour online) The spin-pressure distribution $n_s$, divided into its $z$-projection $n_s^{z}$ and $xy$-projection part $n_s^{xy}$ as defined in \Eq{spinspec_split}, at four different time steps. 
    The time slices as well as the numerical parameters correspond to \Fig{mdens}, and averages are taken over 100 runs. 
    In the intermediate stage both quantities show the scaling behaviour $n_s(k) \sim k^{-3}$ at intermediate momenta (Panel (a)) which is expected for kink solutions. 
    During the following time evolution (Panels (b)--(d)) the scaling of $n_s^z$ is overtaken by a thermal distribution up to a small intermediate momentum region. 
    In contrast, the distribution of $n_s^{xy}$ develops the bimodal scaling form which is characteristic for vortical flows in two dimensions with an IR scaling exponent $\zeta = 4$ (Panel (d)). See the main text for further details.}
    \label{fig:mspecasy_spin}
    \end{figure}
%===============================================================================================================
%
%-------------------------------------------------------------------------------------------------
\subsubsection{Hydrodynamic and spin-fluid decomposition}
It is again instructive to decompose the spectrum according to the different parts of the radial kinetic energy density, \Eq{radenergyspec}. 
The result is shown in \Fig{mspecasy_decomp} for the latest time step in Figs.~\ref{fig:mdens} and \ref{fig:mspecasym}. 
Similar to the immiscible regime and in agreement with simulations for one-component gases the quantum pressure contribution $n_q(k)$ and the compressible energy contribution $n_c(k)$ are negligible for low momenta.
They contribute mainly to the UV modes, with  $\zeta = 2$, reflecting a thermal distribution. 
The incompressible part of the spectrum $n_i(k)$ on the contrary displays the strong-wave turbulent scaling distribution $n_i(k) \sim k^{-4}$ in the IR due to the incompressible flow around vortices and is negligible for the UV region. 
In contrast to the immiscible regime, the spin-pressure contribution $n_s(k)$ follows here also the $k^{-4}$ scaling for low-momentum modes, identical to that of the full single-particle spectrum. 
However, in a small intermediate momentum region the spin-pressure contribution deviates from the $k^{-4}$ scaling, below the scale where the thermal distribution $n_s(k) \sim k^{-2}$ sets in.\\ 

Since the spin order parameter of the model \eq{energy} is not entirely $O(3)$ symmetric for $\alpha \neq 1$ a better insight in the behaviour of the related correlation function $\mathcal{S}$ is gained by splitting it into a part parallel to and orthogonal to the $z$-projection of the spin, $\mathcal{S}^z$ and $\mathcal{S}^{xy}$, respectively. 
Note that for $\rho_T = \mathrm{const.}$ the spin-pressure distribution and the correlation function are proportional (see \Eq{strucfunc}). 
Therefore we can define also a splitting of the spin pressure according to the symmetry of the spin order parameter,
%
%===============================================================================================================
	\begin{subequations}
	\label{eq:spinspec_split}
	  \begin{align}
	\nonumber      n_{s}^z(k) &= \frac{\rho_T}{2}  \int \! \mathrm{d}\Omega_k \, \langle S^z(-\vector{k}) S^z(\vector{k}) \rangle \,\\ 
			         &= \frac{1}{2} \int \! \mathrm{d}\Omega_k \, \langle \lvert\vector{w}_{s}^z(k)\rvert^2 \rangle \label{subeq:spinspec_split-1} \,
	%\\
	  \end{align}
	  \begin{align}
	\nonumber      n_{s}^{xy}(k) &= \frac{\rho_T}{2}  \int \! \mathrm{d}\Omega_k \! \sum_{a \in \{x,y\}} \langle S^a(-\vector{k}) S^a(\vector{k}) \rangle \,\\
				     &= \frac{1}{2} \int \! \mathrm{d}\Omega_k \! \sum_{a \in \{x,y\}} \langle  \lvert\vector{w}_{s}^a(k)\rvert^2  \rangle \,, \label{subeq:spinspec_split-2}  
	  \end{align}
	\end{subequations}
%=============================================================================================================== 
such that the distribution $n_s^z$ is only sensitive to the $S^z$ polarisation while $n_s^{xy}$ reflects spin correlation in the $xy$ plane of the spin state space. 
Due to the strong polarisation in the immiscible regime the influence of the latter onto the particle spectrum or the spin pressure is negligible whereas in the miscible regime both parts are present and each undergo a different time evolution. 
\Fig{mspecasy_spin} depicts the spin pressure contribution separated into $n_s^z$ and $n_s^{xy}$ according to \Eq{spinspec_split}, at four different times corresponding to the snapshots shown in \Fig{mdens}. 
In the intermediate stage (see panel a in \Fig{mspecasy_spin}) both distributions follow a scaling behaviour $n_s \sim k^{-3}$ in a momentum range $k_{\mathrm{min}} < k < k_{\mathrm{max}}$ which is typical for randomly distributed (pseudo-)kinks. 
Since kinks in $S^z$ are not favoured by the ground state properties of the miscible regime spin domains decay here instead of growing and the scale $k_{\mathrm{min}}$ does not move towards lower momenta. 
On the other hand, thermalisation sets in for the UV modes and consequently the scaling regime with $n_s(k) \sim k^{-3}$ vanishes for $n_s^{z}$ during the thermalisation process up to a small intermediate range of momenta (see Panel (d) in \Fig{mspecasy_spin}). 
Note that vortices lead to $S^z$-polarised patches in their core area and thus preserve the $\sim k^{-3}$ scaling of $n_s^z$. 
However, the distribution $n_s^{xy}$ inherits its dynamical evolution during the late stage from the dynamics of vortices. 
Identifying the $xy$-projection of the spin with a complex field $S^+ = S^x + \i S^y$ the relation $n_s^{xy}(k) \sim \int \! \mathrm{d}\Omega_k \, \langle S^{+\ast}(k)S^+(k)\rangle$ holds and $S^+ = ({\sqrt{\rho_1\rho_2}}/{\rho_T})\exp\{\i\theta_r\}$. 
Thus, for the later stages of the evolution, $n_s^{xy}$ behaves like the correlation function of a two-dimensional complex order parameter which contains randomly distributed vortex anti-vortex pairs. 
Their dynamics imprints the IR scaling behaviour $n_s \sim k^{-4}$ onto the distribution $n_s^{xy}$ (Panel (d) in \Fig{mspecasy_spin}), in complete analogy to the vortex binding-unbinding dynamics in a one-component gas in two dimensions.\\
%
%
%===============================================================================================================
\subsection{Dynamics at the transition point}
\label{subsec:TransPoint}         
%
%===============================================================================================================
    \begin{figure}[!t]
    \includegraphics[width=0.49\textwidth]{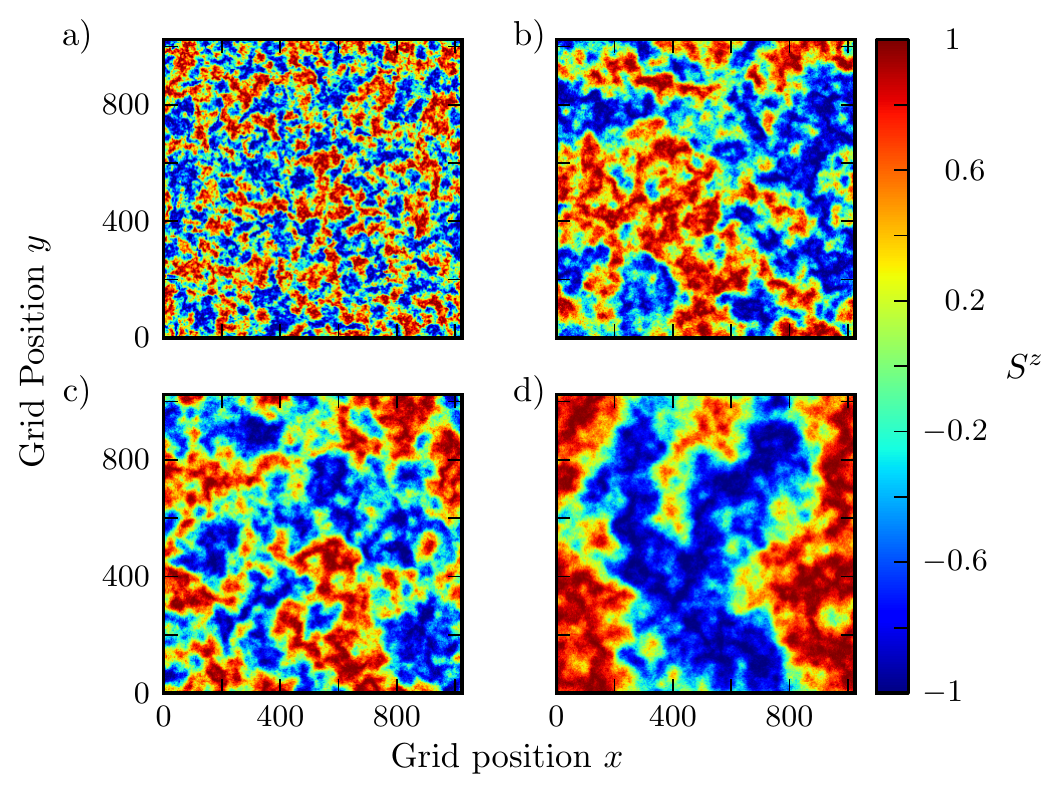}
\includegraphics[width=0.49\textwidth]{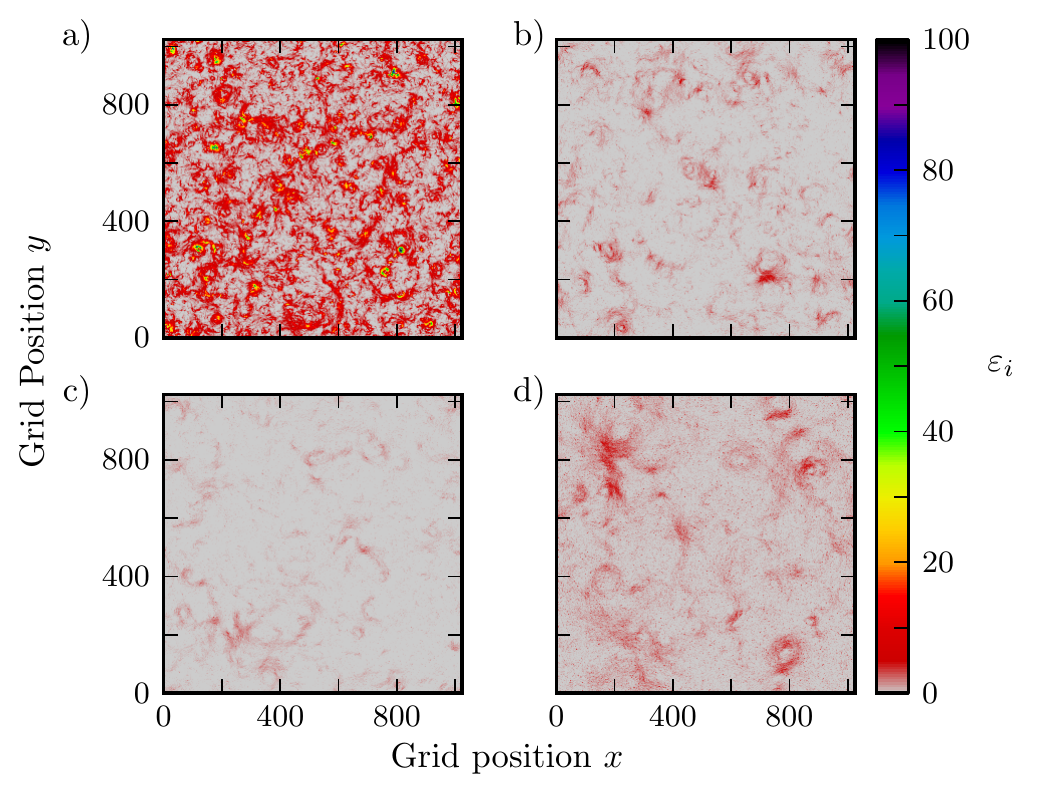}
    \caption{(Colour online) The time evolution of the spin density $S^z(\vector{x})$ (upper half) and of the incompressible kinetic energy density $\epsilon_i = \lvert\vector{w}_i(\vector{x})\rvert^2/2$ (lower half).
    The ratio $\alpha$ of couplings is tuned to the transition point between the miscible and immiscible regimes ($\alpha = 1$). 
    Besides this, numerical parameters and depicted times are the same as for \Fig{mdens}. 
    Note that the incompressible energy density has been amplified by a factor of $6$ for $t = 500000$ in Panel (d). 
    Similar to the situation deep in the miscible regime (compare with \Fig{mdens}), after the onset stage of the counter-superflow instability $S^z$-polarised spin configurations emerge at intermediate times which are accompanied by uniform distributions of incompressible energy $\epsilon_i$ (Panel (a)). 
    However, here the pseudo-domain structure of $S^z$ undergoes a coarse-graining process similar to the situation in the immiscible regime (compare with \Fig{immdens}), instead of decaying in time.
    }
    \label{fig:mdens_sym}
    \end{figure}
%===============================================================================================================
%===============================================================================================================
    \begin{figure}[!t]
    \includegraphics[width=0.49\textwidth]{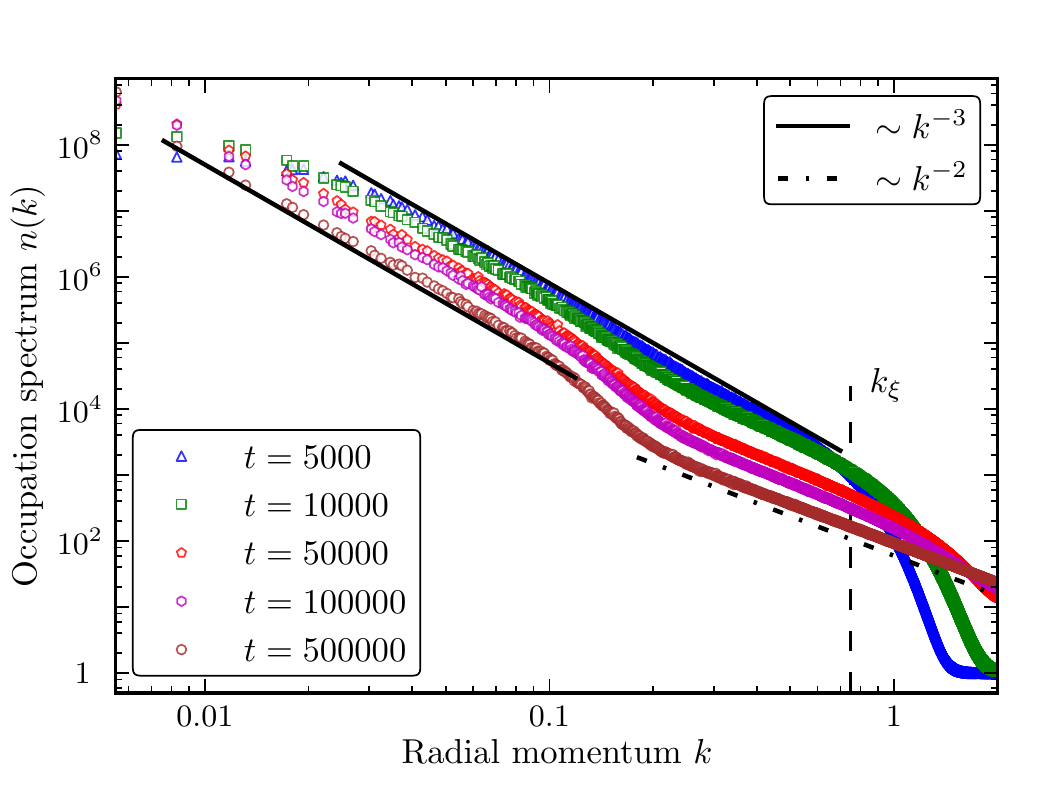}
    \caption{(Colour online) The single-particle momentum distribution as defined in \Eq{sinpartspec}, at different grid times, at the transition point $\alpha = 1$. 
Numerical parameters correspond to those in \Fig{mdens_sym}, and the spectra are averaged over 100 runs. 
Similar to the miscible regime randomly distributed and orientated pseudo-kinks at intermediate times imply a domain-wall scaling $n(k) \sim k^{-3}$ over a wide range of momenta. 
However, during the following evolution this scaling is retained and the region shifted towards the lower momentum modes while thermalisation of high momentum modes sets in. 
At the latest time step we find a momentum distribution with an almost stationary scaling $n(k) \sim k^{-3}$ in the IR.}
    \label{fig:mspecsym}
    \end{figure}
%===============================================================================================================
%
%===============================================================================================================
    \begin{figure}[!t]
    \includegraphics[width=0.49\textwidth]{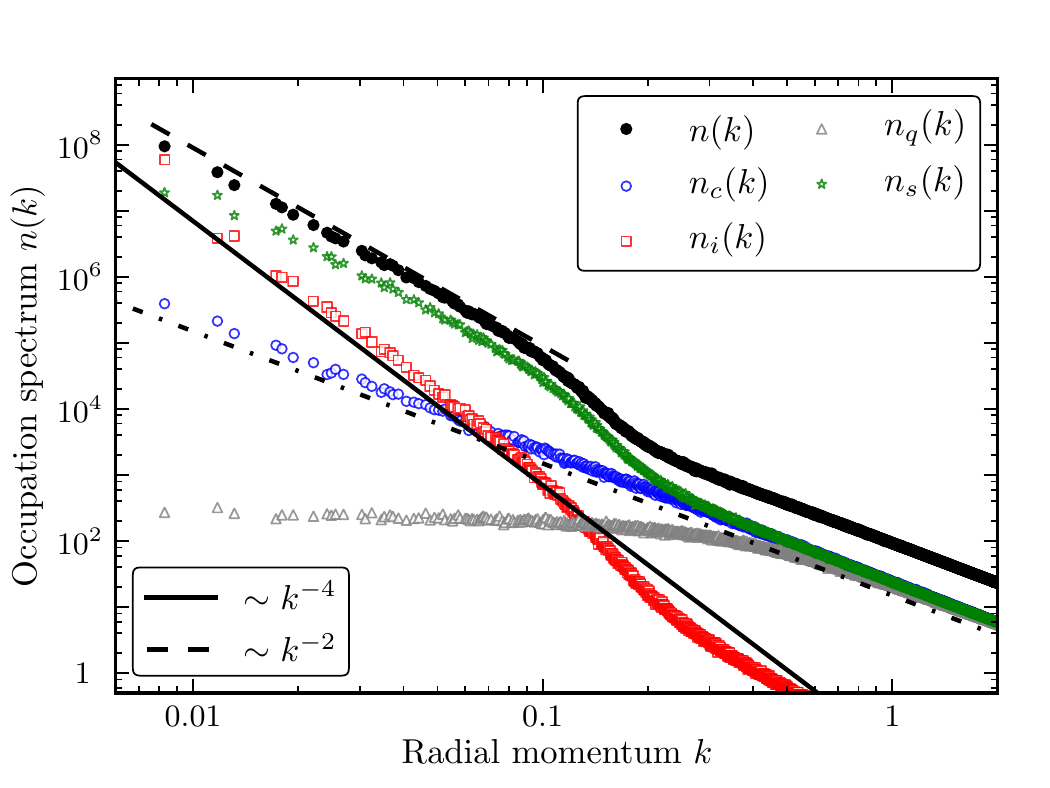}
    \caption{(Colour online) The decomposition of the single-particle spectrum as defined in \Eq{radenergyspec}, in comparison with the full spectrum at grid time $t = 500000$.
     Numerical parameters correspond to those in \Fig{mdens_sym}. 
     Averages are taken over 100 runs. 
     The black dots show the full spectrum which is also displayed in \Fig{mspecsym} while open coloured symbols refer to the incompressible and compressible parts of the hydrodynamic kinetic energy, $n_i$ (red squares) and $n_c$ (blue circles), respectively, 
to the quantum pressure part $n_q$ (grey triangles) and the spin pressure part $n_s$ (green stars). 
In contrast to the situation deep in the miscible and immiscible regimes the spin-pressure contribution $n_s$ dominates the spectrum of excitations for low momenta with a domain-wall scaling similar to that of $n$, $n_s(k) \sim k^{-3}$ for $k_{d} < k < k_{\xi}$.
In this regime, incompressible energy excitations are found to be less relevant. See the main text for further details.}
    \label{fig:mspecsy_decomp}
    \end{figure}
%===============================================================================================================
%===============================================================================================================
    \begin{figure}[!t]
    \includegraphics[width=0.49\textwidth]{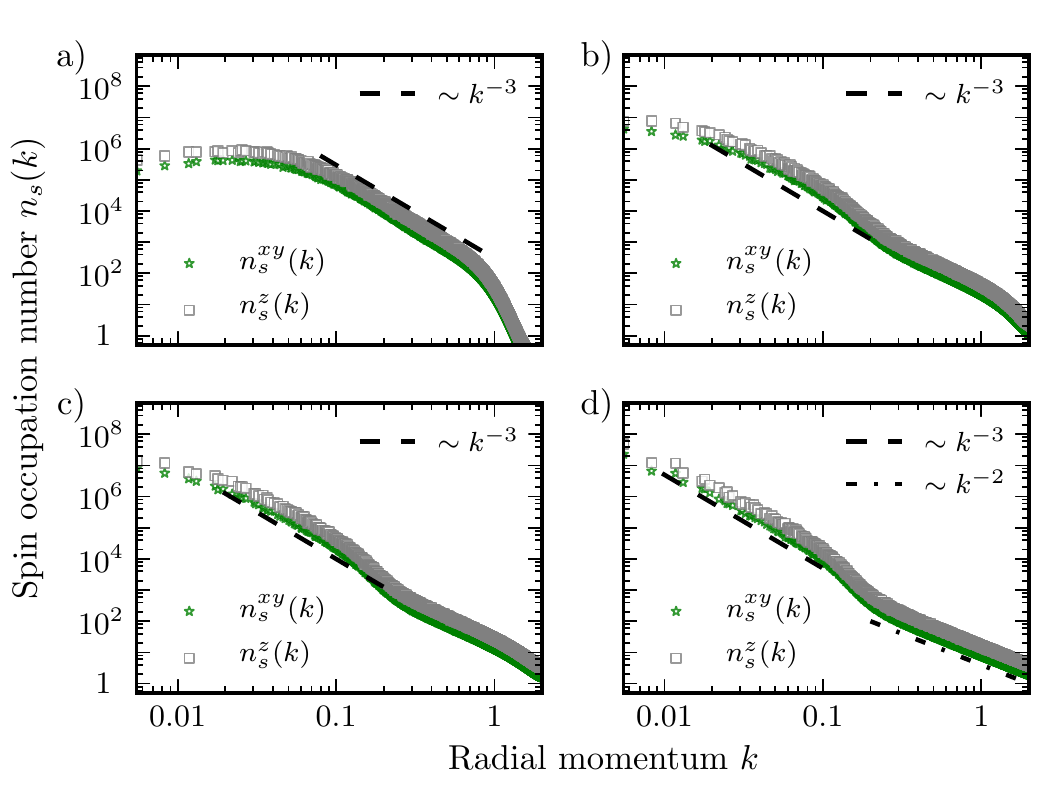}
    \caption{(Colour online) The spin-pressure distribution $n_s$ separated into its $z$-projection $n_s^{z}$ and $xy$-projection $n_s^{xy}$ as defined in \Eq{spinspec_split}, at four different time steps. 
    Those as well as the numerical parameters correspond to \Fig{mdens_sym}, and averages are taken over 100 runs. 
    For the intermediate and late times of evolution depicted here the $xy$-projection follows the form of the $z$-projection, thereby developing the same domain-wall scaling in the IR, $n_s^{xy}(k) \sim n_s^z\sim  k^{-3}$ for $k_{d} < k < k_{\xi}$. 
    Confer the main text for further details.}
    \label{fig:mspecsy_spin}
    \end{figure}
%===================================================
%
%-------------------------------------------------------------------------------------------------
\subsubsection{Symmetry considerations}
In view of our findings in the immiscible and miscible regimes discussed in the previous sections it is particularly interesting to study  the dynamic evolution of the two-component Bose gas directly at the transition point, i.e., for $\alpha = 1$, where no state of mixture is preferred over the other from the perspective of their potential energies. 
This particular choice of $\alpha$ restores full $SU(2)$ symmetry in the model \eq{action}.
Hence, taking the field-theoretic perspective, we simulate the dynamic evolution of a non-relativistic $\phi^4$ theory with $N=2$ field components, for which in turn predictions drawn within a ${1}/{N}$-expansion \cite{Scheppach:2009wu} should apply even better. 
As in \Sect{MiscRegime} the counter-superflow-instability (CSI) mechanism is  used to drive the spin system away from equilibrium during the early evolution. 
Therefore, we use the same ensemble of initial field configurations as for the simulations presented in \subsect{MiscRegime}.
For comparability with the miscible regime, the same numerical parameters including the initial counterflow velocity $v=0.7\xi^{-1}$ are chosen, although at the transition point any initial counterflow is unstable, i.e., $v_{\mathrm{crit}} = 0$.

%-------------------------------------------------------------------------------------------------
\subsubsection{Spatial evolution}
Starting from this numerical preset a typical time evolution of a single realisation is shown in \Fig{mdens_sym}.
We concentrate again on the spin density $S^z$ and the distribution of hydrodynamic incompressible energy $\epsilon_i$. 
In complete analogy to the evolution in the miscible regime CSI-generated pseudo kinks in $S^z$ develop into an isotropic structure of domain-like patches at intermediate times.
This is accompanied by an isotropic distribution of incompressible energy (see Panel (a) in \Fig{mdens_sym} and compare to panel (a) in \Fig{mdens}). 
However, in the following stage of evolution these structures in $S^z$ undergo a coarse-graining process, i.e., they grow, very similar to the dynamic evolution in the immiscible regime, while incompressible energy is decaying (Panels (b--c) in \Fig{mdens_sym}). 
We emphasise that even though configurations with large patches of $S^z \neq 0$ are energetically favourable for $\alpha = 1$, the curvature $\lvert\nabla \vector{S}\rvert^2 \neq 0$ at interfaces between them tends to level out. 
Hence, also at the transition point the pseudo-domain structure in $S^z$ would be expected to decay back to the unpolarised state, especially since interfaces are not protected by means of  topological constraints. 
However, most remarkably, the spin system does not reach or even come close to the unpolarised state over the simulated time span, in contrast to the situation for a miscible system (see \Fig{mdens_sym}d and compare with \Fig{mdens}d). 
In addition, there is no nucleation of stable one-component vortices with coherent hydrodynamic vortical flow around them during the early or intermediate stages.
  
This qualitative observation of stable domain-like structures that slow down the evolution of the spin system towards the unpolarised state and therefore also the thermalisation process of the whole system is supported on a quantitative level by the time evolution of the single-particle momentum distribution, depicted in \Fig{mspecsym}. 
At intermediate times after the onset stage of the CSI the spectrum has developed a scaling $n(k) \sim k^{-3}$ in a momentum region $k_{d} < k < k_{\xi}$ consistent with randomly distributed and randomly orientated pseudo-kinks. 
During the late stages the scale connected to the characterisic size of the pseudo-domain structure $k_d$ moves towards the IR, thus signalling a coarse-graining of the structure. 
In contrast to the immiscible case, here the intrinsic scale of interfaces $\xi$ is not fixed by a topological solution.
This circumstance allows for a reduction of interface energy by broadening, and therefore the scale $k_{\xi}$ moves also towards lower momenta. 
However, the defect-induced scaling $n(k) \sim k^{-3}$ is retained between those two scales during the whole simulated time span. 
At the latest time step in \Fig{mspecsym}, a slowly evolving spectrum has formed that shows thermal scaling $n(k) \sim k^{-2}$ for high momenta and $n(k) \sim k^{-\zeta^{IR}}$ in the low-$k$ regime, with $\zeta^{IR} \simeq 3$. 
Thus, at the transition point $\alpha=1$ we find a deviation in the IR scaling exponent from the field-theoretic prediction in $d=2$ dimensions, $\zeta^{IR} = 4$.
We attribute this to the absence of dominant vortical excitations.      
%

%-------------------------------------------------------------------------------------------------
\subsubsection{Hydrodynamic and spin-fluid decomposition}
In \Fig{mspecsy_decomp}, a decomposition of the single-particle spectrum shown in \Fig{mspecsym} into hydrodynamic and spin-related degrees of freedom is shown for the latest time step. 
We find that as before the quantum-pressure and compressible-energy contributions are negligible for low momenta and contribute equally to the thermal tail for high momenta, together with the spin excitations. 
In congruence with our findings for the IR dynamics at the transition point so far, the spin-pressure component $n_s$ gives the dominant contribution for low momenta, thereby reflecting the defect scaling of $n$, $n_s(k) \sim k^{-3}$ for $k_{d} < k < k_{\xi}$. 
In contrast to the situation deep in the miscible and immiscible regimes here the incompressible energy excitations are below the spin-pressure contribution within a wide range of IR momenta. 
Nevertheless, the incompressible occupation spectrum develops a scaling $n_i(k) \sim k^{-4}$ in the IR, which hints at vortical excitations or flow hidden under the spin polarisation. 
We remark that due to the enlarged ground state manifold of the $SU(2)$ symmetric model vortices in a single component loose their topological stability since they can simply `unwind'. 
Nevertheless,  animations \cite{TwoComponentBoseVideos} of the incompressible energy density show that vortices indeed are created in our simulations but persist only on very short timescales before they decay. 
This is enough to produce a visible IR scaling behaviour of $n(k) \sim k^{-4}$ in the spectrum of incompressible excitations.

Focusing on the main contribution to the spectrum of excitations, \Fig{mspecsy_spin} depicts the time evolution of the two projections $n_s^z$ and $n_s^{xy}$ of the spin pressure. 
At the transition point, the $xy$-projection follows the functional form of the $S^z$-projection during the intermediate and late stages of evolution, in accordance with the $O(3)$ symmetry in the spin system for $\alpha = 1$. 
Thus, the $xy$-projection develops also the defect scaling in the IR, and consequently the corresponding spin densities $S^x$ and $S^y$ develop 
domain-like structures as well. 
%
%
%== Summary ============================================================================
\section{Conclusions}
\label{sec:summary}
We have discussed the non-equilibrium dynamical evolution of a near-degenerate two-component Bose gas towards equilibration, following dynamical instabilities induced by the initial conditions. 
The main results concern the investigation of non-thermal-fixed-point behaviour across the miscible-immiscible transition. 
Note, that the underlying concept of \mbox{(non-)}topological defects determining bulk features of correlation functions in far-from-equilibrium situations is very general. 
It is easily imaginable that multi-component field theories with more than two components show behaviour similar to the one outlined here. 
This requires the generation of (quasi-)topo\-logical configurations far from thermal equilibrium and their slow decay, going together 
with an increase of coherence and defect separation~\cite{Schole:2012kt}. 
Under these conditions, an inverse particle cascade is generated, and the associated power laws can be found from the scaling properties of the respective single defects.
New interesting features that are readily accessible in experiment are expected for ultra cold spinor gases, with spin 1/2 \cite{Timmermans1998a,Hall1998a, Hall1998b,Nicklas2011a,Guzman2011a} or higher~\cite{Stenger1999a, Miesner1999a, Sadler2006a, Vengalattore2008a, Kawaguchi2010a,Ueda2012a, Fujimoto2012a}. 
The transition between different types of transient non-equilibrium order can be controlled by changing the symmetry properties of the Hamiltonian and thus topology and local conservation laws of the system. 
This offers interesting prospects for far-from-equilibrium dynamical transitions in very different areas of physics.

Much work has been done recently concerning the question whether a system on the way to thermalization or general equilibration can approach, underway, non-equilibrium states with characteristics of thermal or generalised Gibbs ensembles, so-called prethermalisation \cite{Berges:2004ce,Gring2011a, Barnett2011a, Kollar2011a, Guzman2011a}, and non-thermal fixed points \cite{Berges:2008wm}.
Prethermalisation is also discussed in the context of the question under which conditions a closed interacting quantum system can thermalise, cf., e.g., \cite{Polkovnikov2011a}.
Prethermalisation in the sense of Refs.~\cite{Berges:2004ce,Gring2011a} can by definition be seen as a mean-field (Gaussian) non-thermal fixed point reached through dephasing of quasiparticle modes occupied in the initial state.
In contrast, the non-thermal fixed points studied here represent stationary configurations where interactions play a vital role.
These fixed points are reached not long after the initial mean-field dephasing period through scattering processes which redistribute particles to form the non-thermal algebraic momentum spectra.
It is expected that within a general renormalisation-group theory of far-from-equilibrium critical phenomena \cite{Gasenzer:2008zz}, prethermalisation phenomena and non-thermal fixed points are treated on equal footings.

%==============================================================================
%
\textit{Acknowledgements}.
We thank J. Berges, S. Diehl, S. Erne, P. Kevrekidis, L. McLerran, E. Nicklas, M. K. Oberthaler, J. M. Pawlowski, J. Schole, D. Sexty, and C. Wetterich for discussions. 
This work was supported by Deutsche Forschungsgemeinschaft (GA677/7,8), the University of Heidelberg (CQD), and the Helmholtz Association (HA216/EMMI).

% Create the reference section using BibTeX:
%\bibliography{./Bibliography/Master}

\end{document}